\pdfoutput=1
\documentclass[american,11pt,a4paper,DIV=11]{scrartcl}
\usepackage[T1]{fontenc}
\usepackage[utf8]{inputenc}
\usepackage{lmodern}
\usepackage{float}
\usepackage{url}
\usepackage{amsmath}
\usepackage{graphicx}
\usepackage{amssymb}
\usepackage{nicefrac}

\makeatletter

\newcommand{\strong}[1]{\textbf{#1}}
\floatstyle{ruled}
\newfloat{algorithm}{tbp}{loa}
\floatname{algorithm}{Algorithm}

\usepackage{tikz}
\usepackage{clrscode3e}

\usepackage[unicode=true, 
            bookmarks=true,
            bookmarksnumbered=true,
            bookmarksopen=true,
            breaklinks=true,
            pdfborder={0 0 0}, 
            colorlinks=false   
]{hyperref}

\usepackage{url}

\hypersetup{pdftitle={Average Case Analysis of Java 7’s Dual Pivot Quicksort},
 pdfauthor={Sebastian Wild and Markus E. Nebel},
 pdfsubject={},
 pdfkeywords={Quicksort, Dual Pivot, Average Case Analysis}}

\def\SKIP#1{}

\newdimen\makeboxdimen
\newcommand\makeboxlike[3][l]{%
\setbox0=\hbox{#2}%
\global\makeboxdimen=\wd0%
\setbox1=\hbox{\makebox[\makeboxdimen][#1]{%
\makebox[0pt][#1]{#3}%
}}%
\ht1=\ht0%
\dp1=\dp0%
\box1%
}

\def\EndIf{\End\li\kw{end if} }
\def\EndWhile{\End\li\kw{end while} }

\newcommand\E{\mathop{\mbox{$\mathbb{E}$}}\nolimits}

\makeatother

\usepackage{babel}

\usepackage{tocbibind}
\addtokomafont{caption}{\sffamily\small}
\addtokomafont{captionlabel}{\sffamily\textbf}
\setcapmargin{1em}

\usepackage{amsthm}
\theoremstyle{definition}
\newtheorem{property}{Property}

\newtheorem{example}{Example}

\theoremstyle{plain}

\newtheorem{lemma}{Lemma}

\renewcommand\proof[1][]{%
    \ifthenelse{\equal{#1}{}}{%
        \textit{Proof: }%
    }{%
        \textit{Proof #1:}%
    }%
}

\usepackage{scrpage2}

\clearscrheadings
\automark[section]{section}

\title{Average Case Analysis of \\
Java\,7's Dual Pivot Quicksort\thanks{This research was supported by DFG grant NE 1379/3-1.}}

\author{%
	Sebastian Wild${}^\dagger$
	\and 
	Markus E. Nebel\thanks{%
		Fachbereich Informatik, Technische Universität Kaiserslautern, 
		\texttt{\{wild,nebel\}@cs.uni-kl.de}
	}%
}

\begin{document}

\maketitle

\vspace*{-5ex}

\begin{abstract}
\noindent\paragraph{Abstract.}
Recently, a new Quicksort variant due to Yaroslavskiy was chosen as
standard sorting method for Oracle's Java 7 runtime library. The decision
for the change was based on empirical studies showing that on average,
the new algorithm is faster than the formerly used classic Quicksort.
Surprisingly, the improvement was achieved by using a dual pivot approach,
an idea that was considered not promising by several theoretical studies
in the past. In this paper, we identify the reason for this unexpected
success. Moreover, we present the first precise average case analysis
of the new algorithm showing e.\,g.\ that a random permutation of
length $n$ is sorted using $1.9n\ln n-2.46n+\mathcal{O}(\ln n)$
key comparisons and $0.6n\ln n+0.08n+\mathcal{O}(\ln n)$ swaps.
\end{abstract}
\global\long\def\like#1#2{\makeboxlike[c]{\ensuremath{#1}}{\ensuremath{#2}}}
\global\long\def\likel#1#2{\makeboxlike[l]{\ensuremath{#1}}{\ensuremath{#2}}}

\section{Introduction}

Due to its efficiency in the average, Quicksort has been used for
decades as general purpose sorting method in many domains, e.\,g.\ in
the C and Java standard libraries or as UNIX's system sort. Since
its publication in the early 1960s by Hoare \cite{Hoare1962}, classic
Quicksort (Algorithm~\ref{alg:Classic-Quicksort}) has been intensively
studied and many modifications were suggested to improve it even further,
one of them being the following: Instead of partitioning the input
file into two subfiles separated by a single pivot, we can create
$s$ partitions out of $s-1$ pivots.

Sedgewick considered the case $s=3$ in his PhD thesis \cite{Sedgewick1975}.
He proposed and analyzed the implementation given in Algorithm~\ref{alg:Dual-Pivot-Quicksort-Sedgewick}.
However, this dual pivot Quicksort variant turns out to be clearly
inferior to the much simpler classic algorithm. Later, Hennequin studied
the comparison costs for any constant $s$ in his PhD thesis~\cite{hennequin1991analyse},
but even for arbitrary $s\ge3$, he found no improvements that would
compensate for the much more complicated partitioning step.%
\footnote{When $s$ depends on $n$, we basically get the Samplesort algorithm
from~\cite{Frazer1970}. \cite{Sanders2004}, \cite{Leischner2009}
or \cite{Blelloch} show that Samplesort can beat Quicksort if hardware
features are exploited. \cite{Sedgewick1975}~even shows that Samplesort
is asymptotically optimal with respect to comparisons. Yet, due to
its inherent intricacies, it has not been used much in practice.%
} These negative results may have discouraged further research along
these lines.

\begin{algorithm}
\vspace{-1.5ex}\begin{codebox}
\Procname{$\proc{Quicksort}(A,\id{left},\id{right})$}
\zi \Comment Sort the array $A$ in index range $\id{left},\dots,\id{right}$. We assume a sentinel $A[0]=-\infty$.
\li \If $\id{right} - \id{left} \ge 1$
\li \Then 
 		$p\gets A[\id{right}]$ \>\>\>\>\Comment Choose rightmost element as pivot
\li		$i\gets \id{left}-1$;\>\>\>$j\gets \id{right}$
\li 	\kw{do}\Do
\li			\kw{do} $i\gets i+1$ \While $A[i] < p$ \kw{end while}
\li			\kw{do} $j\gets j-1$ \While $A[j] > p$ \kw{end while}
\li			\If $j>i$ \kw{then} Swap $A[i]$ and $A[j]$ \kw{end if} 
		\End
\li		\While $j > i$
\li		Swap $A[i]$ and $A[\id{right}]$ \>\>\>\>\>\Comment Move pivot to final position
\li		$\proc{Quicksort}(A,
				\makeboxlike[c]{$j+1$}{$\id{left}$},
				\makeboxlike[c]{$j+1$}{$i-1$}
			)$
\li		$\proc{Quicksort}(A,
				\makeboxlike[c]{$j+1$}{$i+1$},
				\makeboxlike[c]{$j+1$}{$\id{right}$}
			)$
	\EndIf
\end{codebox}

\vspace{-1.5ex}

\caption{~Implementation of classic Quicksort\label{alg:Classic-Quicksort}
\strut as given in~\cite{Sedgewick1978} (see \cite{Sedgewick1975},
\cite{Sedgewick1977a} and~\cite{Sedgewick1977} for detailed analyses).\protect \\
Two pointers $i$ and $j$ scan the array from left and right until
they hit an element that does not belong in their current subfiles.
Then the elements $A[i]$ and $A[j]$ are exchanged. This {}``crossing
pointers'' technique is due to Hoare \cite{Hoare1961b}, \cite{Hoare1962}.\strut}

\end{algorithm}

\begin{algorithm}
\vspace{-1.5ex}
\begin{codebox}
\Procname{$\proc{DualPivotQuicksortSedgewick}(A,\id{left},\id{right})$}
\zi \Comment Sort the array $A$ in index range $\id{left},\dots,\id{right}$.
\li \If $\id{right} - \id{left} \ge 1$
\li \Then 
 		$i\gets \id{left}$;~~$i_1\gets \id{left}$;~~$j\gets \id{right}$;~~$j_1\gets \id{right}$;~~$p\gets A[\id{left}]$;~~$q\gets A[\id{right}]$
\li		\If $p > q$ \label{lin:sedgewick-comp-0}
		\kw{then} 
			Swap $p$ and $q$ \label{lin:sedgewick-swap-0}
		\kw{end if}
\li		\While $\mathit{true}$
\li 	\Do
			$i\gets i+1$
\li			\While $A[i] \le q$ \label{lin:sedgewick-comp-1}
\li			\Do
				\If $i \ge j$ \kw{then} \kw{break} outer while \kw{end if} \>\>\>\>\>\>\Comment pointers have crossed

\li				\If $A[i] < p$ \kw{then} $A[i_1]\gets A[i]$; $i_1\gets i_1+1$; $A[i]\gets A[i_1]$ \kw{end if}
					\label{lin:sedgewick-comp-2}
					\label{lin:sedgewick-swap-1}
\li				$i\gets i+1$
			\EndWhile
\li			$j\gets j-1$
\li			\While $A[j] \ge p$ \label{lin:sedgewick-comp-1'}
\li			\Do
				\If $A[j] > q$ \kw{then} $A[j_1]\gets A[j]$; $j_1\gets j_1-1$; $A[j]\gets A[j_1]$ \kw{end if}
					\label{lin:sedgewick-comp-2'}
					\label{lin:sedgewick-swap-1'}
\li				\If $i \ge j$ \kw{then} \kw{break} outer while \kw{end if} \>\>\>\>\>\>\Comment pointers have crossed
\li				$j\gets j-1$
			\EndWhile
\li			$A[i_1]\gets A[j]$; \>\>\>$A[j_1]\gets A[i]$ \label{lin:sedgewick-swap-2}
\li			$i_1\gets i_1+1$;   \>\>\>$j_1\gets j_1-1$
\li			$A[i]\gets A[i_1]$; \>\>\>$A[j]\gets A[j_1]$
		\EndWhile
\li		$A[i_1]\gets p$;~~$A[j_1]\gets q$ \label{lin:sedgewick-swap-3}
\li		$\proc{DualPivotQuicksortSedgewick}(A,
				\makeboxlike[c]{$j_1+1$}{$\id{left}$},
				\makeboxlike[c]{$j_1+1$}{$i_1-1$}
			)$
\li		$\proc{DualPivotQuicksortSedgewick}(A,
				\makeboxlike[c]{$j_1+1$}{$i_1+1$},
				\makeboxlike[c]{$j_1+1$}{$j_1-1$}
			)$
\li		$\proc{DualPivotQuicksortSedgewick}(A,
				\makeboxlike[c]{$j_1+1$}{$j_1+1$},
				\makeboxlike[c]{$j_1+1$}{$\id{right}$}
			)$
	\EndIf
\end{codebox}

\vspace{-1.5ex}

\caption{Dual Pivot Quicksort \strut with \label{alg:Dual-Pivot-Quicksort-Sedgewick}Sedgewick's
partitioning as proposed in~\cite{Sedgewick1975} (Program 5.1).
This is an equivalent Java-like adaption of the original ALGOL-style
program.\strut }

\end{algorithm}

\begin{algorithm}
\vspace{-1.5ex}
\begin{codebox}
\Procname{$\proc{DualPivotQuicksortYaroslavskiy}(A,\id{left},\id{right})$}
\zi \Comment Sort the array $A$ in index range $\id{left},\dots,\id{right}$.
\li \If $\id{right} - \id{left} \ge 1$
\li \Then 
		$p\gets A[\id{left}]$; \>\>\> $q\gets A[\id{right}]$
\li		\If $p > q$ \label{lin:yaroslavskiy-comp-0}
		\kw{then} 
			Swap $p$ and $q$ \label{lin:yaroslavskiy-swap-0}
		\kw{end if}
\li 	$\ell\gets \id{left}+1$; \>\>\> $g\gets \id{right}-1$; \>\>\>$\quad$ $k\gets \ell$
\li		\While $k\le g$
\li 	\Do
			\If $A[k] < p$ \label{lin:yaroslavskiy-comp-1}
\li			\Then
				Swap $A[k]$ and $A[\ell]$ \label{lin:yaroslavskiy-swap-1}
\li				$\ell\gets \ell+1$
\li			\Else 
\li				\If $A[k] > q$ \label{lin:yaroslavskiy-comp-2}
\li				\Then
					\While $A[g] > q$ and $k<g$ \kw{do} $g\gets g-1$ \kw{end while} \label{lin:yaroslavskiy-comp-3}
\li					Swap $A[k]$ and $A[g]$ \label{lin:yaroslavskiy-swap-2}
\li					$g\gets g-1$ \label{lin:yaroslavskiy-g--}
\li					\If $A[k] < p$ \label{lin:yaroslavskiy-comp-4}
\li					\Then
						Swap $A[k]$ and $A[\ell]$ \label{lin:yaroslavskiy-swap-3}
\li						$\ell\gets \ell+1$
					\EndIf
				\EndIf
			\EndIf
\li			$k\gets k+1$
		\EndWhile \label{lin:yaroslavskiy-end-while}
\li		$\ell\gets \ell-1$; \>\>\>$g\gets g+1$
\li		Swap $A[\id{left}]$ and $A[\ell]$ \label{lin:yaroslavskiy-swap-4}
			\>\>\>\>\>\Comment Bring pivots to final position
\li		Swap $A[\id{right}]$ and $A[g]$ \label{lin:yaroslavskiy-swap-5}
\li		$\proc{DualPivotQuicksortYaroslavskiy}(A,
				\makeboxlike[c]{$g+1$}{$\id{left}$},
				\makeboxlike[c]{$g+1$}{$\ell-1$}
			)$
\li		$\proc{DualPivotQuicksortYaroslavskiy}(A,
				\makeboxlike[c]{$g+1$}{$\ell+1$},
				\makeboxlike[c]{$g+1$}{$g-1$}
			)$
\li		$\proc{DualPivotQuicksortYaroslavskiy}(A,
				\makeboxlike[c]{$g+1$}{$g+1$},
				\makeboxlike[c]{$g+1$}{$\id{right}$}
			)$
	\EndIf
\end{codebox}

\vspace{-1.5ex}

\caption{Dual Pivot Quicksort \strut with \label{alg:Dual-Pivot-Quicksort-Yaroslavskiy}Yaroslavskiy's
partitioning method}

\end{algorithm}

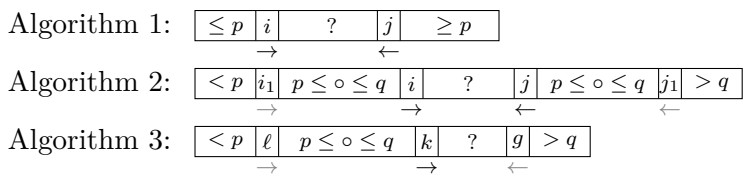
\begin{figure}[tbh]
\vspace*{1ex}
\mbox{}\hfill\parbox{0.7\linewidth}{%
	\quad{}Algorithm~\ref{alg:Classic-Quicksort}: \begin{scriptsize}
	\begin{tikzpicture}[scale=0.4,baseline=(ref.south)]
	\draw (0,0) -- ++(10,0) -- ++(0,1) -- ++(-10,0) -- cycle;
	\node at (1,0.5) {$\le p$};
	\node at (8.5,0.5) {$\ge p$};
	\node[inner sep=0pt] (ref) at (4.5,0.5) {?};
	\draw (2,1) -- ++(0,-1);
	\draw (2.75,1) -- ++ (0,-1);
	\node at (2.375,0.5) {$i$};
	\node[below] at (2.375,0) {$\rightarrow$};
	\draw (6,1) -- ++(0,-1);
	\draw (6.75,1) -- ++ (0,-1);
	\node at (6.375,0.5) {$j$};
	\node[below] at (6.375,0) {$\leftarrow$};
	\end{tikzpicture}
	\end{scriptsize}

	\quad{}Algorithm~\ref{alg:Dual-Pivot-Quicksort-Sedgewick}: \begin{scriptsize}
	\begin{tikzpicture}[scale=0.4,baseline=(ref.south)]
	\draw (0,0) -- ++(18,0) -- ++(0,1) -- ++(-18,0) -- cycle;
	\node at (1,0.5) {$< p$};
	\draw (2,1) -- ++(0,-1);
	\draw (2.75,1) -- ++ (0,-1);
	\node at (2.375,0.5) {$i_1$};
	\node[below,gray] at (2.375,0) {$\rightarrow$};

	\node at (4.75,0.5) {$p\le \circ\le q$};
	\draw (6.75,1) -- ++(0,-1);
	\draw (7.5,1) -- ++ (0,-1);
	\node at (7.125,0.5) {$i$};
	\node[below] at (7.125,0) {$\rightarrow$};

	\node[inner sep=0pt] (ref) at (9,0.5) {?};

	\draw (10.5,1) -- ++(0,-1);
	\node at (10.875,0.5) {$j$};
	\node[below] at (10.875,0) {$\leftarrow$};

	\node at (13.25,0.5) {$p\le \circ\le q$};
	\draw (11.25,1) -- ++(0,-1);
	\draw (15.25,1) -- ++(0,-1);

	\node at (15.625,0.5) {$j_1$};
	\node[below,gray] at (15.625,0) {$\leftarrow$};
	\draw (16,1) -- ++ (0,-1);
	\node at (17,0.5) {$>q$};
	\end{tikzpicture}
	\end{scriptsize}

	\quad{}Algorithm~\ref{alg:Dual-Pivot-Quicksort-Yaroslavskiy}: \begin{scriptsize}
	\begin{tikzpicture}[scale=0.4,baseline=(ref.south)]
	\draw (0,0) -- ++(13,0) -- ++(0,1) -- ++(-13,0) -- cycle;
	\node at (1,0.5) {$< p$};
	\draw (2,1) -- ++(0,-1);
	\draw (2.75,1) -- ++ (0,-1);
	\node at (2.375,0.5) {$\ell$};
	\node[below,gray] at (2.375,0) {$\rightarrow$};

	\node at (12,0.5) {$> q$};
	\draw (10.25,1) -- ++(0,-1);
	\draw (11,1) -- ++ (0,-1);
	\node at (10.625,0.5) {$g$};
	\node[below,gray] at (10.625,0) {$\leftarrow$};

	\node at (5,0.5) {$p\le \circ\le q$};
	\draw (7.25,1) -- ++(0,-1);
	\draw (8,1) -- ++ (0,-1);
	\node at (7.625,0.5) {$k$};
	\node[below] at (7.625,0) {$\rightarrow$};

	\node[inner sep=0pt] (ref) at (9.125,0.5) {?};

	\end{tikzpicture}
	\end{scriptsize}
}\hfill\mbox{}

\caption{Comparison of the \label{fig:partition-schemes}partitioning schemes
of the three Quicksort variants discussed in this paper. The pictures
show the invariant maintained in partitioning.}

\end{figure}

Recently, however, Yaroslavskiy proposed the new dual pivot Quicksort
implementation as given in Algorithm~\ref{alg:Dual-Pivot-Quicksort-Yaroslavskiy}
at the Java core library mailing list%
\footnote{The discussion is archived at \url{http://permalink.gmane.org/gmane.comp.java.openjdk.core-libs.devel/2628}.%
}. He initiated a discussion claiming his new algorithm to be superior
to the runtime library's sorting method at that time: the widely used
and carefully tuned variant of classic Quicksort from~\cite{Bentley1993}.
Indeed, Yaroslavskiy's Quicksort has been chosen as the new default
sorting algorithm in Oracle's Java 7 runtime library after extensive
empirical performance tests.

In light of the results on multi-pivot Quicksort mentioned above,
this is quite surprising and asks for explanation. Accordingly, since
the new dual pivot Quicksort variant has not been analyzed in detail,
yet%
\footnote{Note that the results presented in \url{http://iaroslavski.narod.ru/quicksort/DualPivotQuicksort.pdf}
provide wrong constants and thus are insufficient for our needs.%
}, corresponding average case results will be proven in this paper.
Our analysis reveals the reason why dual pivot Quicksort can indeed
outperform the classic algorithm and why the partitioning method of
Algorithm~\ref{alg:Dual-Pivot-Quicksort-Sedgewick} is suboptimal.
It turns out that Yaroslavskiy's partitioning method is able to take
advantage of certain asymmetries in the outcomes of key comparisons.
Algorithm~\ref{alg:Dual-Pivot-Quicksort-Sedgewick} fails to utilize
them, even though being based on the same abstract algorithmic idea.

\section{Results}

In this paper, we give the first precise average case analysis of
Yaroslavskiy's dual pivot Quicksort (Algorithm~\ref{alg:Dual-Pivot-Quicksort-Yaroslavskiy}),
the new default sorting method in Oracle's Java~7 runtime library.
Using these original results, we compare the algorithm to existing
Quicksort variants: The classic Quicksort (Algorithm~\ref{alg:Classic-Quicksort})
and a dual pivot Quicksort as proposed by Sedgewick in~\cite{Sedgewick1975}
(Algorithm~\ref{alg:Dual-Pivot-Quicksort-Sedgewick}).

\begin{table}
\caption{Exact expected \label{tab:results}number of comparisons and swaps
of the three Quicksort variants in the random permutation model. The
results for Algorithm~\ref{alg:Classic-Quicksort} are taken from
\cite[p. 334]{Sedgewick1977} (for $M=1$). $\mathcal{H}_{n}=\sum_{i=1}^{n}\tfrac{1}{i}$
is the $n$th harmonic number, which is asymptotically $\mathcal{H}_{n}=\ln n+0.577216\ldots+\mathcal{O}(n^{-1})$
as $n\to\infty$.}

\vspace*{.5ex}

\small
\mbox{}\hfill%
\begin{tabular}{lll}
\hline\noalign{\smallskip} & Comparisons & Swaps\tabularnewline
\noalign{\smallskip}\hline\noalign{\smallskip}Classic Quicksort~ & $2(n+1)\mathcal{H}_{n+1}-\frac{8}{3}(n+1)$ & $\frac{1}{3}(n+1)\mathcal{H}_{n+1}-\frac{7}{9}(n+1)+\frac{1}{2}$\tabularnewline
(Algorithm~\ref{alg:Classic-Quicksort}) & $\approx2n\ln n-1.51n+\mathcal{O}(\ln n)$ & $\approx0.33n\ln n-0.58n+\mathcal{O}(\ln n)$\tabularnewline
\noalign{\medskip}Sedgewick & $\frac{32}{15}(n+1)\mathcal{H}_{n+1}-\frac{856}{225}(n+1)+\frac{3}{2}$ & $\frac{4}{5}(n+1)\mathcal{H}_{n+1}-\frac{19}{25}(n+1)-\frac{1}{4}$\tabularnewline
(Algorithm~\ref{alg:Dual-Pivot-Quicksort-Sedgewick}) & $\approx2.13n\ln n-2.57n+\mathcal{O}(\ln n)$ & $\approx0.8n\ln n-0.30n+\mathcal{O}(\ln n)$\tabularnewline
\noalign{\medskip}Yaroslavskiy & $\frac{19}{10}(n+1)\mathcal{H}_{n+1}-\frac{711}{200}(n+1)+\frac{3}{2}$~~ & $\frac{3}{5}(n+1)\mathcal{H}_{n+1}-\frac{27}{100}(n+1)-\frac{7}{12}$\tabularnewline
(Algorithm~\ref{alg:Dual-Pivot-Quicksort-Yaroslavskiy}) & $\approx1.9n\ln n-2.46n+\mathcal{O}(\ln n)$ & $\approx0.6n\ln n+0.08n+\mathcal{O}(\ln n)$\tabularnewline
\end{tabular}%
\hfill\mbox{}
\end{table}

Table~\ref{tab:results} shows formulæ for the expected number of
key comparisons and swaps for all three algorithms. In terms of comparisons,
the new dual pivot Quicksort by Yaroslavskiy is best. However, it
needs more swaps, so whether it can outperform the classic Quicksort,
depends on the relative runtime contribution of swaps and comparisons,
which in turn differ from machine to machine. Section~\ref{sec:Running-Times}
shows some running times, where indeed Algorithm~\ref{alg:Dual-Pivot-Quicksort-Yaroslavskiy}
was fastest.

Remarkably, the new algorithm is significantly better than Sedgewick's
dual pivot Quicksort in both measures. Given that Algorithms~\ref{alg:Dual-Pivot-Quicksort-Sedgewick}
and~\ref{alg:Dual-Pivot-Quicksort-Yaroslavskiy} are based on the
same algorithmic idea, the considerable difference in costs is surprising.
The explanation of the superiority of Yaroslavskiy's variant is a
major discovery of this paper. Hence, we first give a qualitative
teaser of it. Afterwards, Section~\ref{sec:Analysis} gives a thorough
analysis, making the arguments precise.

\subsection{The Superiority of Yaroslavskiy's Partitioning Method}

Let $p<q$ be the two pivots. For partitioning, we need to determine
for every $x\notin\{p,q\}$ whether $x<p$, $p<x<q$ or $q<x$ holds
by comparing $x$ to $p$ \emph{and/or} $q$. Assume, we first compare
$x$ to $p$, then averaging over all possible values for $p$, $q$
and $x$, there is a $\nicefrac{1}{3}$ chance that $x<p$ -- in which
case we are done. Otherwise, we still need to compare $x$ and $q$.
The expected number of comparisons for one element is therefore $\nicefrac{1}{3}\cdot1+\nicefrac{2}{3}\cdot2=\nicefrac{5}{3}$.
For a partitioning step with $n$ elements including pivots $p$ and
$q$, this amounts to $\nicefrac{5}{3}\cdot(n-2)$ comparisons in
expectation.

In the random permutation model, knowledge about an element $y\ne x$
does not tell us whether $x<p$, $p<x<q$ or $q<x$ holds. Hence,
one could think that any partitioning method should need at least
$\nicefrac{5}{3}\cdot(n-2)$ comparisons in expectation. But this
is not the case.

The reason is the independence assumption above, which only holds
true for algorithms that do comparisons at exactly \emph{one location in the code}.
But Algorithms~\ref{alg:Dual-Pivot-Quicksort-Sedgewick} and~\ref{alg:Dual-Pivot-Quicksort-Yaroslavskiy}
have several compare-instructions at different locations, and how
often those are reached \emph{depends} on the pivots $p$ and $q$.
Now of course, the number of elements smaller, between and larger
$p$ and $q$, directly depends on $p$ and $q$, as well! So if a
comparison is executed often if $p$ is large, it is clever to first
check $x<p$ there: The comparison is done more often than on average
if and only if the probability for $x<p$ is larger than on average.
Therefore, the expected number of comparisons can drop below the {}``lower
bound'' $\nicefrac{5}{3}$ for this element!

And this is exactly, where Algorithms~\ref{alg:Dual-Pivot-Quicksort-Sedgewick}
and~\ref{alg:Dual-Pivot-Quicksort-Yaroslavskiy} differ: Yaroslavskiy's
partitioning always evaluates the {}``better'' comparison first,
whereas in Sedgewick's dual pivot Quicksort this is not the case.
In Section~\ref{sub:Superiority-of-Yaroslavskiy}, we will give this
a more quantitative meaning based on our analysis.

\section{Average Case Analysis of Dual Pivot Quicksort\label{sec:Analysis}}

We assume input sequences to be random permutations, i.\,e.\ each
permutation~$\pi$ of elements $\{1,\dots,n\}$ occurs with probability
$\nicefrac{1}{n!}$. The first and last elements are chosen as pivots;
let the smaller one be $p$, the larger one $q$.

Note that all Quicksort variants in this paper fulfill the following
property:
\begin{property}
Every key comparison involves a pivot element of the current partitioning
step.\label{pro:Only-comparisons-with-pivots}
\end{property}

\subsection{Solution to the Dual Pivot Quicksort Recurrence\label{sub:Solution-to-the-Recurrence}}

In~\cite{hennequin1989combinatorial}, Hennequin shows that Property~\ref{pro:Only-comparisons-with-pivots}
is a sufficient criterion for \emph{preserving randomness} in subfiles,
i.\,e.\ if the whole array is a (uniformly chosen) random permutation
of its elements, so are the subproblems Quicksort is recursively invoked
on. This allows us to set up a recurrence relation for the expected
costs, as it ensures that all partitioning steps of a subarray of
size $k$ have the same expected costs as the initial partitioning
step for a random permutation of size~$k$.

The expected costs $C_{n}$ for sorting a random permutation of length
$n$ by any dual pivot Quicksort with Property~\ref{pro:Only-comparisons-with-pivots}
satisfy the following recurrence relation:\begin{align*}
C_{n} & =\sum_{1\le p<q\le n}\Pr[\mbox{pivots }(p,q)]\cdot\left(\mbox{partitioning costs}+\mbox{recursive costs}\right)\\
 & =\sum_{1\le p<q\le n}\frac{2}{n(n-1)}\left(\mbox{partitioning costs}+C_{p-1}+C_{q-p-1}+C_{n-q}\right)\,,\end{align*}
for~$n\ge3$ with base cases $C_{0}=C_{1}=0$ and $C_{2}=d$.%
\footnote{$d$ can easily be determined manually: For Algorithm~\ref{alg:Dual-Pivot-Quicksort-Yaroslavskiy},
it is $1$ for comparisons and $\frac{5}{2}$ for swaps and for Algorithm~\ref{alg:Dual-Pivot-Quicksort-Sedgewick}
we have $d=2$ for comparisons and $d=\frac{5}{2}$ for swaps.%
}

We confine ourselves to linear expected partitioning costs $a(n+1)+b$,
where $a$ and \textbf{$b$} are constants depending on the kind of
costs we analyze. The recurrence relation can then be solved by standard
techniques -- the detailed calculations can be found in Appendix~\ref{sec:Derivation-Recurrence}.
The closed form for $C_{n}$ is \[
C_{n}=\tfrac{6}{5}a\cdot(n+1)\left(\mathcal{H}_{n+1}-\tfrac{1}{5}\right)+\bigl(-\tfrac{3}{2}a+\tfrac{3}{10}b+\tfrac{1}{10}d\,\bigr)\cdot(n+1)-\tfrac{1}{2}b\,,\]
which is valid for $n\ge4$ with $\mathcal{H}_{n}=\sum_{i=1}^{n}\frac{1}{i}$
the $n$th harmonic number.

\subsection{Costs of One Partitioning Step}

In this section, we analyze the expected number of swaps and comparisons
used in the first partitioning step on a random permutation of $\{1,\dots,n\}$.
The results are summarized in Table~\ref{tab:Expected-partition-costs}.
To state the proofs, we need to introduce some notation.%

\begin{table}
\centering{}\caption{Expected costs of the first partitioning step for the two dual pivot
Quicksort \label{tab:Expected-partition-costs}variants on a random
permutation of length $n$ (for $n\ge3$)}

\vspace*{1ex}

\small
\begin{tabular}{lll}
\hline\noalign{\smallskip} & Comparisons & Swaps\tabularnewline
\noalign{\smallskip}\hline\noalign{\medskip}Sedgewick  & $\frac{16}{9}(n+1)-3-\frac{2}{3}\frac{1}{n(n-1)}$~~~ & $\frac{2}{3}(n+1)+\frac{1}{2}$\tabularnewline
(Algorithm~\ref{alg:Dual-Pivot-Quicksort-Sedgewick})~~ &  & \tabularnewline
\noalign{\medskip}Yaroslavskiy & $\frac{19}{12}(n+1)-3$ & $\frac{1}{2}(n+1)+\frac{7}{6}$\tabularnewline
(Algorithm~\ref{alg:Dual-Pivot-Quicksort-Yaroslavskiy}) &  & \tabularnewline
\end{tabular}
\end{table}

\global\long\def\values#1{#1}
\global\long\def\valuesets#1{#1}
\global\long\def\positionsets#1{\mathcal{#1}}

\subsubsection{Notation}

Let $\valuesets S$ be the set of all elements smaller than both pivots,
$\valuesets M$ those in the middle and $\valuesets L$ the large
ones, i.\,e.\ \[
\valuesets S:=\{1,\dots,p-1\},\quad\valuesets M:=\{p+1,\dots,q-1\},\quad\valuesets L:=\{q+1,\dots,n\}\;.\]
Then, by Property~\ref{pro:Only-comparisons-with-pivots} the algorithm
cannot distinguish $x\in C$ from $y\in C$ for any $C\in\{\valuesets S,\valuesets M,\valuesets L\}$.
Hence, for analyzing partitioning costs, we replace all non-pivot
elements by $\values s$, $\values m$ or $\values l$ when they are
elements of $\valuesets S$, $\valuesets M$ or $\valuesets L$, respectively.
Obviously, all possible results of a partitioning step correspond
to the same word $\values{s\cdots s\, p\, m\cdots m\, q\, l\cdots l}$.
The following example will demonstrate these definitions.\smallskip{}

\begin{example}
\noindent \quad{}%
\parbox[t]{0.4\columnwidth}{%
Example permutation before \ldots{}\\
\def\n#1{\makebox[12pt][c]{$\values{#1}$}}
\def\l#1{\makebox[12pt][c]{#1}}
{\footnotesize\n p}\n ~\n ~\n ~\n ~\n ~\n ~\n ~{\footnotesize\n q} \\
\l{\fbox{$2$}}\n 4\n 7\n 8\n 1\n 6\n 9\n 3\l{\fbox{$5$}} \\
\n p\n m\n l\n l\n s\n l\n l\n m\n q %
}\qquad{}%
\parbox[t]{0.4\columnwidth}{%
\ldots{} and after partitioning.\\
~\\
\noindent
\def\n#1{\makebox[12pt][c]{$\values{#1}$}}
\def\l#1{\makebox[12pt][c]{#1}}
\n 1\l{\fbox{$2$}}\n 4\n 3\l{\fbox{$5$}}\n 6\n 9\n 8\n 7 \\
\n s\n p\n m\n m\n q\n l\n l\n l\n l %
}\medskip{}

\end{example}
\noindent  Next, we define position sets $\positionsets S$, $\positionsets M$
and $\positionsets L$ as follows:

\noindent %
\parbox[c]{0.45\columnwidth}{%
\begin{align*}
\positionsets S & :=\{2,\dots,p\},\\
\positionsets M & :=\{p+1,\dots,q-1\},\\
\positionsets L & :=\{q,\dots,n-1\}\;.\end{align*}
}%
\parbox[c]{0.45\columnwidth}{%
in the example:\\
~\\
\def\n#1{\makebox[12pt][c]{$#1$}}
\def\l#1{\makebox[12pt][c]{#1}}
\def\p#1{\makebox[12pt][c]{\footnotesize{$\positionsets{#1}$}}}
\n ~\p S\p M\p M\p L\p L\p L\p L \n ~ \\
\l{\fbox{$2$}}\n 4\n 7\n 8\n 1\n 6\n 9\n 3\l{\fbox{$5$}} \\
\tiny\strut\textsl{\l 1\l 2\l 3\l 4\l 5\l 6\l 7\l 8\l 9}%
}

\global\long\def\numberat#1#2{\values{#1}\,@\,\positionsets{#2}}
\global\long\def\expnumberat#1#2{\E\left[\numberat{#1}{#2}\right]}
\global\long\def\condexpnumberat#1#2{\E\left[\numberat{#1}{#2}\,|\, p,q\right]}

\noindent Now, we can formulate the main quantities occurring in
the analysis below: For a given permutation, $c\in\{\values s,\values m,\values l\}$
and a set of positions $\positionsets P\subset\{1,\dots,n\}$, we
write $\numberat cP$ for the number of $\values c$-type elements
occurring \emph{at positions} in $\positionsets P$ of the permutation.
In our last example, $\positionsets M=\{3,4\}$ holds. At these positions,
we find elements $7$ and $8$ (before partitioning), both belonging
to $\valuesets L$. Thus, $\numberat lM=2$, whereas $\numberat sM=\numberat mM=0$.

Now consider a \emph{random} permutation. Then $\numberat cP$ becomes
a random variable. In the analysis, we will encounter the conditional
expectation of $\numberat cP$ \emph{given} that the random permutation
induces the pivots $p$ and $q$, i.\,e.\ the first and last element
of the permutation are $p$ and $q$ \emph{or} $q$ and $p$, respectively.
We abbreviate this quantity as $\condexpnumberat cP$. As the number
$\#c$ of $\values c$-type elements only depends on the pivots, not
on the permutation itself, $\#c$ is a fully determined constant in
$\condexpnumberat cP$. Hence, given pivots $p$ and $q$, $\numberat cP$
is a hypergeometrically distributed random variable: For the $\values c$-type
elements, we draw their $\#c$ positions out of $n-2$ possible positions
via sampling without replacement. Drawing a position in $\positionsets P$
is a {}`success', a position not in $\positionsets P$ is a {}`failure'. 

Accordingly, $\condexpnumberat cP$ can be expressed as the mean of
this hypergeometric distribution: $\condexpnumberat cP=\#c\cdot\frac{|\positionsets P|}{n-2}$.
By the law of total expectation, we finally have\begin{align*}
\expnumberat cP & =\sum_{1\le p<q\le n}\condexpnumberat cP\cdot\Pr[\mbox{pivots }(p,q)]\\
 & =\frac{2}{n(n-1)}\sum_{1\le p<q\le n}\#c\cdot\frac{|\positionsets P|}{n-2}\;.\end{align*}

\subsubsection{Comparisons in Algorithm~\ref{alg:Dual-Pivot-Quicksort-Yaroslavskiy}}

Algorithm~\ref{alg:Dual-Pivot-Quicksort-Yaroslavskiy} contains five
places where key comparisons are used, namely in lines~\ref{lin:yaroslavskiy-comp-0},
\ref{lin:yaroslavskiy-comp-1}, \ref{lin:yaroslavskiy-comp-2}, \ref{lin:yaroslavskiy-comp-3}
and~\ref{lin:yaroslavskiy-comp-4}. Line~\ref{lin:yaroslavskiy-comp-0}
compares the two pivots and is executed exactly once. Line~\ref{lin:yaroslavskiy-comp-1}
is executed once per value for $k$ except for the last increment,
where we leave the loop before the comparison is done. Similarly,
line~\ref{lin:yaroslavskiy-comp-3} is run once for every value of
$g$ except for the last one.

The comparison in line~\ref{lin:yaroslavskiy-comp-2} can only be
reached, when line~\ref{lin:yaroslavskiy-comp-1} made the {}`else'-branch
apply. Hence, line~\ref{lin:yaroslavskiy-comp-2} causes as many
comparisons as $k$ attains values with $A[k]\ge p$. Similarly, line~\ref{lin:yaroslavskiy-comp-4}
is executed once for all values of $g$ where $A[g]\le q$.%
\footnote{Line \ref{lin:yaroslavskiy-swap-2} just swapped $A[k]$ and $A[g]$.
So even though line~\ref{lin:yaroslavskiy-comp-4} literally says
“$A[k]<p$”, this comparison actually refers to an element first reached
as $A[g]$.\label{fn:Why-A[k]-is-A[g]}%
}

At the end, $q$ gets swapped to position $g$ (line~\ref{lin:yaroslavskiy-swap-5}).
Hence we must have $g=q$ there. Accordingly, $g$ attains values
$\positionsets G=\{n-1,n-2,\dots,q\}=\positionsets L$ at line~\ref{lin:yaroslavskiy-comp-3}.
We always leave the outer while loop with $k=g+1$ or $k=g+2$. In
both cases, $k$ (at least) attains values $\positionsets K=\{2,\dots,q-1\}=\positionsets S\cup\positionsets M$
in line~\ref{lin:yaroslavskiy-comp-3}. The case “$k=g+2$” introduces
an additional term of $3\cdot\tfrac{n-q}{n-2}$; see Appendix~\ref{sec:Corner-Case}
for the detailed discussion.

Summing up all contributions yields the conditional expectation $c_{n}^{p,q}$
of the number of comparisons needed in the first partitioning step
for a random permutation, given it implies pivots $p$ and $q$:\begin{align*}
c_{n}^{p,q} & =1+|\positionsets K|+|\positionsets G|+\bigl(\condexpnumberat mK+\condexpnumberat lK\bigr)\\
 & \hphantom{\mbox{}=\mbox{}1+|\positionsets K|+|\positionsets G|}+\bigl(\condexpnumberat sG+\condexpnumberat mG\bigr)\\
 & \hphantom{\mbox{}=\mbox{}1+|\positionsets K|+|\positionsets G|}+3\cdot\tfrac{n-q}{n-2}\\
 & =n-1+\bigl((q-p-1)+(n-q)\bigr)\frac{q-2}{n-2}\\
 & \hphantom{\mbox{}=\mbox{}n-1}+\bigl((p-1)+(q-p-1)\bigr)\frac{n-q}{n-2}\\
 & \hphantom{\mbox{}=\mbox{}n-1}+3\cdot\tfrac{n-q}{n-2}\\
 & =n-1+\bigl(n-p-1\bigr)\frac{q-2}{n-2}+\bigl(q+1\bigr)\frac{n-q}{n-2}\;.\end{align*}
Now, by the law of total expectation, the expected number of comparisons
in the first partitioning step for a random permutation of $\{1,\dots,n\}$
is\begin{align*}
c_{n}:=\E c_{n}^{p,q} & =\tfrac{2}{n(n-1)}\sum_{p=1}^{n-1}\sum_{q=p+1}^{n}c_{n}^{p,q}\\
 & =n-1+\tfrac{2}{n(n-1)(n-2)}\sum_{p=1}^{n-1}(n-p-1)\sum_{q=p+1}^{n}(q-2)\\
 & \hphantom{\mbox{}=n-1}+\tfrac{2}{n(n-1)(n-2)}\sum_{q=2}^{n}(n-q)(q+1)\sum_{p=1}^{q-1}1\\
 & =n-1+\left(\tfrac{5}{12}(n+1)-\tfrac{4}{3}\right)+\tfrac{1}{6}(n+3)\;=\;\tfrac{19}{12}(n+1)-3\;.\end{align*}

\subsubsection{Swaps in Algorithm~\ref{alg:Dual-Pivot-Quicksort-Yaroslavskiy}}

Swaps happen in Algorithm~\ref{alg:Dual-Pivot-Quicksort-Yaroslavskiy}
in lines~\ref{lin:yaroslavskiy-swap-0}, \ref{lin:yaroslavskiy-swap-1},
\ref{lin:yaroslavskiy-swap-2}, \ref{lin:yaroslavskiy-swap-3}, \ref{lin:yaroslavskiy-swap-4}
and~\ref{lin:yaroslavskiy-swap-5}. Lines~\ref{lin:yaroslavskiy-swap-4}
and~\ref{lin:yaroslavskiy-swap-5} are both executed exactly once.
Line~\ref{lin:yaroslavskiy-swap-0} once swaps the pivots if needed,
which happens with probability $\nicefrac{1}{2}$. For each value
of $k$ with $A[k]<p$, one swap occurs in line~\ref{lin:yaroslavskiy-swap-1}.
Line~\ref{lin:yaroslavskiy-swap-2} is executed for every value of
$k$ having $A[k]>q$. Finally, line~\ref{lin:yaroslavskiy-swap-3}
is reached for all values of $g$ where $A[g]<p$ (see footnote~\ref{fn:Why-A[k]-is-A[g]}). 

Using the ranges $\positionsets K$ and $\positionsets G$ from above,
we obtain $s_{n}^{p,q}$, the conditional expected number of swaps
for partitioning a random permutation, given pivots $p$ and $q$.
There is an additional contribution of $\frac{n-q}{n-2}$ when $k$
stopps with $k=g+2$ instead of $k=g+1$. As for comparisons, its
detailed discussion is deferred to Appendix~\ref{sec:Corner-Case}.\begin{align*}
s_{n}^{p,q} & =\tfrac{1}{2}+1+1+\condexpnumberat sK+\condexpnumberat lK+\condexpnumberat sG+\tfrac{n-q}{n-2}\\
 & =\tfrac{5}{2}+(p-1)\frac{q-2}{n-2}+(n-q)\frac{q-2}{n-2}+(p-1)\frac{n-q}{n-2}+\tfrac{n-q}{n-2}\\
 & =\tfrac{5}{2}+(n+p-q-1)\frac{q-2}{n-2}+p\cdot\frac{n-q}{n-2}\;.\end{align*}
Averaging over all possible $p$ and $q$ again, we find\begin{align*}
s_{n}:=\E s_{n}^{p,q} & =\tfrac{5}{2}+\tfrac{2}{n(n-1)(n-2)}\sum_{q=2}^{n}(q-2)\sum_{p=1}^{q-1}(n+p-q-1)\\
 & \hphantom{\mbox{}=\tfrac{5}{2}}+\tfrac{2}{n(n-1)(n-2)}\sum_{q=2}^{n}(n-q)\sum_{p=1}^{q-1}p\\
 & =\tfrac{5}{2}+\left(\tfrac{5}{12}(n+1)-\tfrac{4}{3}\right)+\tfrac{1}{12}(n+1)\;=\;\tfrac{1}{2}(n+1)+\tfrac{7}{6}\;.\end{align*}

\subsubsection{Comparisons in Algorithm~\ref{alg:Dual-Pivot-Quicksort-Sedgewick}}

Key comparisons happen in Algorithm~\ref{alg:Dual-Pivot-Quicksort-Sedgewick}
in lines~\ref{lin:sedgewick-comp-0}, \ref{lin:sedgewick-comp-1},
\ref{lin:sedgewick-comp-2}, \ref{lin:sedgewick-comp-1'} and~\ref{lin:sedgewick-comp-2'}.
Lines~\ref{lin:sedgewick-comp-1} and~\ref{lin:sedgewick-comp-1'}
are executed once for every value of $i$ respectively $j$ (without
the initialization values $\mathit{left}$ and $\mathit{right}$ respectively).
Line~\ref{lin:sedgewick-comp-2} is reached for all values of $i$
with $A[i]\le q$ except for the last value. Finally, the comparison
in line~\ref{lin:sedgewick-comp-2'} gets executed for every value
of $j$ having $A[j]\ge p$.

The value-ranges of $i$ and $j$ are $\positionsets I=\{2,\dots,\hat{\imath}\}$
and $\positionsets J=\{n-1,{n-2},\dots,\hat{\imath}\}$ respectively,
where $\hat{\imath}$ depends on the positions of $\values m$-type
elements. So, lines~\ref{lin:sedgewick-comp-1} and~\ref{lin:sedgewick-comp-1'}
together contribute $|\positionsets I|+|\positionsets J|=n-1$ comparisons.
For lines~\ref{lin:sedgewick-comp-2} and~\ref{lin:sedgewick-comp-2'},
we get additionally \[
\bigl(\condexpnumberat s{I'}+\condexpnumberat m{I'}\bigr)+\bigl(\condexpnumberat mJ+\condexpnumberat lJ\bigr)\]
many comparisons (in expectation), where $\positionsets{I'}:=\positionsets I\setminus\hat{\imath}$.
As $i$ and $j$ cannot meet on an $\values m$-type element (both
would not stop), $\numberat m{\mathnormal{\{\hat{\imath}\}}}=0$,
so \[
\condexpnumberat m{I'}+\condexpnumberat mJ=q-p-1\;.\]

Positions of $\values m$-type elements do not contribute to $\numberat s{I'}$
(and $\numberat lJ$) by definition. Hence, it suffices to determine
the number of non-$m$-elements located at positions in~$\positionsets{I'}$.
A glance at Figure~\ref{fig:partition-schemes} suggests to count
non-$\values m$-type elements left of (and including) the last value
of $i_{1}$, which is $p$. So, the first $p-1$ of all $(p-1)+(n-q)$
non-$m$-positions are contained in $\positionsets{I'}\!$, thus $\condexpnumberat s{I'}=(p-1)\frac{p-1}{(p-1)+(n-q)}$.
Similarly, we can show that $\numberat lJ$ is the number of $\values l$-type
elements right of $i_{1}$'s largest value: $\condexpnumberat lJ=(n-q)\frac{n-q}{(p-1)+(n-q)}$.
Summing up all contributions, we get\[
{c'}{}_{n}^{p,q}=n-1+q-p-1+(p-1)\tfrac{p-1}{(p-1)+(n-q)}+(n-q)\tfrac{n-q}{(p-1)+(n-q)}\;.\]
Taking the expectation over all possible pivot values yields \[
c'_{n}=\tfrac{2}{n(n-1)}\sum_{p=1}^{n-1}\sum_{q=p+1}^{n}{c'}_{n}^{p,q}=\tfrac{16}{9}(n+1)-3-\tfrac{2}{3}\tfrac{1}{n(n-1)}\;.\]

This is not a linear function and hence does not directly fit our
solution of the recurrence from Section~\ref{sub:Solution-to-the-Recurrence}.
The exact result given in Table~\ref{tab:results} is easily proven
by induction. Dropping summand $-\tfrac{2}{3}\tfrac{1}{n(n-1)}$ and
inserting the linear part into the recurrence relation, still gives
the correct leading term; in fact, the error is only $\frac{1}{90}(n+1)$.

\subsubsection{Swaps in Algorithm~\ref{alg:Dual-Pivot-Quicksort-Sedgewick}}

The expected number of swaps has already been analyzed in~\cite{Sedgewick1975}.
There, it is shown that Sedgewick's partitioning step needs $\frac{2}{3}(n+1)$
swaps, on average -- excluding the pivot swap in line~\ref{lin:sedgewick-swap-0}.
As we count this swap for Algorithm~\ref{alg:Dual-Pivot-Quicksort-Yaroslavskiy},
we add $\frac{1}{2}$ to the expected value for Algorithm~\ref{alg:Dual-Pivot-Quicksort-Sedgewick},
for consistency.

\subsection{Superiority of Yaroslavskiy's Partitioning Method -- Continued\label{sub:Superiority-of-Yaroslavskiy}}

\begin{table}[tb]
\caption{$\expnumberat cP$ for $c=s,m,l$ and $P=\positionsets S,\positionsets M,\positionsets L$.\label{tab:expnumberat}}

\vspace*{1ex}

\small
\centering\begin{tabular}{llll}
\hline 
\noalign{\vskip\doublerulesep}
 & {\small $\positionsets S$} & {\small $\positionsets M$} & {\small $\positionsets L$}\tabularnewline[\doublerulesep]
\hline
\noalign{\vskip\doublerulesep}
{\small $\values s$} & {\small $\frac{1}{6}(n-1)$} & {\small $\frac{1}{12}(n-3)$} & {\small $\frac{1}{12}(n-3)$}\tabularnewline
\noalign{\vskip\doublerulesep}
{\small $\values m$~~} & {\small $\frac{1}{12}(n-3)$~~} & {\small $\frac{1}{6}(n-1)$~~} & {\small $\frac{1}{12}(n-3)$}\tabularnewline
\noalign{\vskip\doublerulesep}
{\small $\values l$} & {\small $\frac{1}{12}(n-3)$} & {\small $\frac{1}{12}(n-3)$} & {\small $\frac{1}{6}(n-1)$}\tabularnewline
\end{tabular}
\end{table}

In this section, we abbreviate $\expnumberat cP$ by\global\long\def\expnumberat#1#2{E_{#1}^{\positionsets{#2}}}
$\expnumberat cP$ for conciseness. It is quite enlightening to compute
$\expnumberat cP$ for $\values c=\values s,\values m,\values l$
and $\positionsets P=\positionsets S,\positionsets M,\positionsets L$,
see Table~\ref{tab:expnumberat}: There is a remarkable \emph{asymmetry},
e.\,g.\ averaging over all permutations, \emph{more than half }of
all $l$-type elements are located at positions in $\positionsets L$.
Thus, if we \emph{know} we are looking at a position in $\positionsets L$,
it is much more advantageous to first compare with $q$, as with probability
$>\frac{1}{2}$, the element is $>q$. This results in an expected
number of comparisons $<\frac{1}{2}\cdot2+\frac{1}{2}\cdot1=\frac{3}{2}<\frac{5}{3}$.
Line~\ref{lin:yaroslavskiy-comp-3} of Algorithm~\ref{alg:Dual-Pivot-Quicksort-Yaroslavskiy}
is exactly of this type. Hence, Yaroslavskiy's partitioning method
exploits the knowledge about the different position sets comparisons
are reached for. Conversely, lines~\ref{lin:sedgewick-comp-1} and~\ref{lin:sedgewick-comp-1'}
in Algorithm~\ref{alg:Dual-Pivot-Quicksort-Sedgewick} are of the
opposite type: They check the unlikely outcome first.
\pagebreak

We can roughly approximate the expected number of comparisons in Algorithms~\ref{alg:Dual-Pivot-Quicksort-Sedgewick}
and~\ref{alg:Dual-Pivot-Quicksort-Yaroslavskiy} by expressing them
in terms of the quantities from Table~\ref{tab:expnumberat} (using
$\positionsets K=\positionsets S\cup\positionsets M$, $\positionsets G\approx\positionsets L$
and $\expnumberat s{I'}+\expnumberat lJ\approx\expnumberat sS+\expnumberat lL+\expnumberat sM$):\begin{align*}
c'_{n} & =n-1+\like{\bigl(\expnumberat mS+\expnumberat mM+\expnumberat mL\bigr)}{\E\#m}+\like{\bigl(\expnumberat sS+\expnumberat lL+\expnumberat sM\bigr)}{\expnumberat s{I'}+\expnumberat lJ}\\
 & \approx\like{n-1}n+\bigl(\expnumberat mS+\expnumberat mM+\expnumberat mL\bigr)+\bigl(\expnumberat sS+\expnumberat lL+\expnumberat sM\bigr)\\
 & \approx(1+3\cdot\tfrac{1}{12}+3\cdot\tfrac{1}{6})n\;\approx\;1.75n\qquad(\mbox{exact: }1.78n-1.22+o(1))\\
c_{n} & =\vphantom{a^{a^{a^{a}}}}n+\like{\bigl(\expnumberat mS+\expnumberat mM\bigr)}{\expnumberat mK}+\like{\bigl(\expnumberat lS+\expnumberat lM\bigr)}{\expnumberat lK}+\expnumberat sG+\expnumberat mG\\
 & \approx n+\bigl(\expnumberat mS+\expnumberat mM\bigr)+\bigl(\expnumberat lS+\expnumberat lM\bigr)+\expnumberat sL+\expnumberat mL\\
 & \approx(1+5\cdot\tfrac{1}{12}+1\cdot\tfrac{1}{6})n\;\approx\;1.58n\qquad(\mbox{exact: }1.58n-0.75)\end{align*}
Note that both terms involve six {}`$\expnumberat cP$-terms', but
Algorithm~\ref{alg:Dual-Pivot-Quicksort-Sedgewick} has \emph{three}
{}`expensive' terms, whereas Algorithm~\ref{alg:Dual-Pivot-Quicksort-Yaroslavskiy}
only has \emph{one} such term.

\global\long\def\expnumberat#1#2{\E\left[\numberat{#1}{#2}\right]}

\section{Some Running Times\label{sec:Running-Times}}

\begin{figure}
\noindent \begin{centering}
\includegraphics[width=0.66\columnwidth]{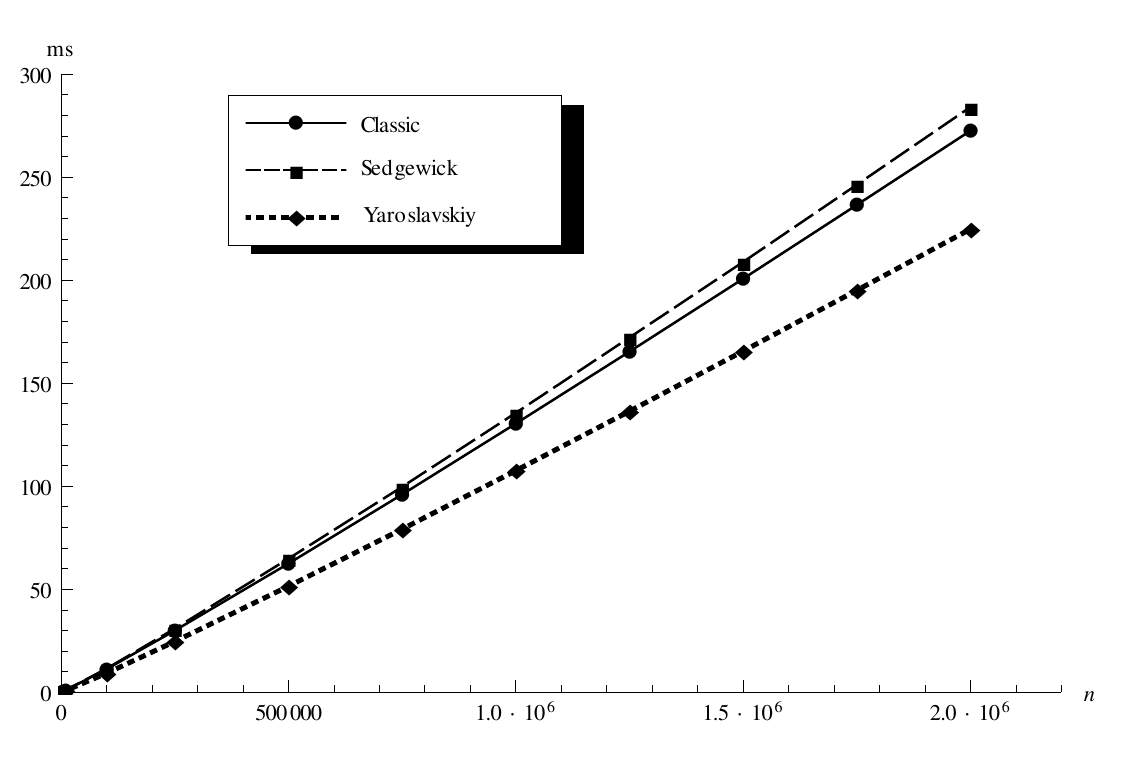}
\par\end{centering}

\caption{Running times of Java \label{fig:Running-times}implementations of
Algorithms~\ref{alg:Classic-Quicksort}, \ref{alg:Dual-Pivot-Quicksort-Sedgewick}
and~\ref{alg:Dual-Pivot-Quicksort-Yaroslavskiy} on an Intel Core
2 Duo P8700 laptop. The plot shows the average running time of 1000
random permutations of each size.}

\end{figure}

Extensive performance tests have already been done for Yaroslavskiy's
dual pivot Quicksort. However, those were based on an optimized implementation
intended for production use. In Figure~\ref{fig:Running-times},
we provide some running times of the basic variants as given in Algorithms~\ref{alg:Classic-Quicksort},
\ref{alg:Dual-Pivot-Quicksort-Sedgewick} and~\ref{alg:Dual-Pivot-Quicksort-Yaroslavskiy}
to directly evaluate the algorithmic ideas, complementing our analysis.

Note: This is not intended to replace a thorough performance study,
but merely to demonstrate that Yaroslavskiy's partitioning method
performs well -- at least on our machine.

\section{Conclusion and Future Work}

Having understood how the new Quicksort saves key comparions, there
are plenty of future research directions. The question if and how
the new Quicksort can compensate for the many extra swaps it needs,
calls for further examination. One might conjecture that comparisons
have a higher runtime impact than swaps. It would be interesting to
see a closer investigation -- empirically or theoretically.

In this paper, we only considered the most basic implementation of
dual pivot Quicksort. Many suggestions to improve the classic algorithm
are also applicable to it. We are currently working on the effect
of selecting the pivot from a larger sample and are keen to see the
performance impacts.

Being intended as a standard sorting method, it is not sufficient
for the new Quicksort to perform well on random permutations. One
also has to take into account other input distributions, most notably
the occurrence of equal keys or biases in the data. This might be
done using Maximum Likelihood Analysis as introduced in~\cite{Laube2010},
which also helped us much in discovering the results of this paper.
Moreover, Yaroslavskiy's partitioning method can be used to improve
Quickselect. Our corresponding results are omitted due to space constraints.

\bibliographystyle{plain}
\begin{footnotesize}
\bibliography{quicksort-refs}
\end{footnotesize}

\pagebreak{}

\appendix

\section{Solution of the Dual Pivot Quicksort Recurrence\label{sec:Derivation-Recurrence}}

The presented analysis is a generalization of the derivation given
by Sedgewick in \cite[p. 156ff]{Sedgewick1975}. In \cite{hennequin1991analyse},
Hennequin gives an alternative approach based on generating functions
that is much more general. Even though the authors consider Hennequin's
method more elegant, we prefer the elementary proof, as it allows
a self-contained presentation.

The expected costs $C_{n}$ for sorting a random permutation of length
$n$ by any dual pivot Quicksort fulfilling Property~\ref{pro:Only-comparisons-with-pivots}
satisfy the following recurrence relation (for~$n\ge2$):\begin{align*}
C_{n} & =\sum_{1\le p<q\le n}\Pr[\mbox{pivots }(p,q)]\cdot\left(\mbox{partitioning costs}+\mbox{recursive costs}\right)\\
 & =\sum_{1\le p<q\le n}\frac{2}{n(n-1)}\left(\mbox{partitioning costs}+C_{p-1}+C_{q-p-1}+C_{n-q}\right)\\
 & =\E\mbox{partitioning costs}+\frac{2}{n(n-1)}\cdot3\sum_{k=0}^{n-2}(n-k-1)C_{k}\;.\end{align*}
(The last equation follows from splitting up the sum and shifting
indices.)\\
As both algorithms skip subfiles of length $\le1$, the base case
is $C_{0}=C_{1}=0$.

We will solve this recurrence relation for linear expected partitioning
costs $a(n+1)+b$, where $a$ and \textbf{$b$} are constants depending
on the kind of costs we analyze. It turns out that the costs for (sub)lists
of length $n=2$ do not fit the linear pattern. Hence, we add $C_{2}=d$
as an additional base case and use the recurrence for $n\ge3$.

We first consider $D_{n}:=\tbinom{n+1}{2}C_{n+1}-\tbinom{n}{2}C_{n}$
to get rid of the factor in the sum: \begin{align*}
D{}_{n} & =\tbinom{n+1}{2}\bigl(a(n+2)+b\bigr)-\tbinom{n}{2}\bigl(a(n+1)+b\bigr)\\
 & \hphantom{\mbox{}=\mbox{}}+\tfrac{(n+1)n}{2}\tfrac{6}{(n+1)n}\sum_{k=0}^{n-1}(n-k)C_{k}-\tfrac{n(n-1)}{2}\tfrac{6}{n(n-1)}\sum_{k=0}^{n-2}(n-k-1)C_{k}\\
 & =3\tbinom{n+1}{2}a+n\cdot b+3\sum_{k=0}^{n-1}C_{k}\;.\qquad\qquad(n\ge3)\end{align*}
The remaining full history recurrence can be solved by taking ordinary
differences $E_{n}:=D_{n+1}-D_{n}=3(n+1)a+b+3C_{n}$ for $n\ge3$.
Using the definition of $E_{n}$ and some tedious, yet elementary
rearrangements we find \[
\left(E_{n}-3C_{n}\right)\bigm/\tbinom{n+2}{2}=C_{n+2}-\tfrac{2n}{n+2}C_{n+1}+\tfrac{n-3}{n+1}C_{n}\;.\]
Considering yet another quantity $F_{n}:=C_{n}-\frac{n-4}{n}\cdot C_{n-1}$,
one easily checks that $F_{n+2}-F_{n+1}=C_{n+2}-\tfrac{2n}{n+2}C_{n+1}+\tfrac{n-3}{n+1}C_{n}$
holds, such that we conclude\[
F_{n+2}-F_{n+1}=\left(E_{n}-3C_{n}\right)\bigm/\tbinom{n+2}{2}=\bigl(3(n+1)a+b\bigr)\bigm/\tbinom{n+2}{2}\;.\qquad(n\ge3)\]
This last equation is now amenable to simple iteration:\begin{align*}
F_{n} & =\sum_{i=5}^{n}\bigl(3(i-1)a+b\bigr)\bigm/\!\tbinom{i}{2}+F_{4}\\
 & =\sum_{i=5}^{n}\frac{3(i-1)a}{\frac{1}{2}i(i-1)}+\sum_{i=5}^{n}\frac{b}{\frac{1}{2}i(i-1)}+F_{4}\\
 & =6a\sum_{i=5}^{n}\tfrac{1}{i}+2b\sum_{i=5}^{n}\left(\tfrac{1}{i-1}-\tfrac{1}{i}\right)+F_{4}\\
 & =6a(\mathcal{H}_{n}-\mathcal{H}_{4})+2b\left(\tfrac{1}{4}-\tfrac{1}{n}\right)+F_{4}\;.\qquad(n\ge5)\end{align*}
($\mathcal{H}_{n}:=\sum_{i=1}^{n}\nicefrac{1}{i}$ is the $n$th harmonic
number.)\\
Plugging in the definition of $F_{n}=C_{n}-\frac{n-4}{n}\cdot C_{n-1}$
yields \[
C_{n}=\tfrac{n-4}{n}\cdot C_{n-1}\;+\;6a(\mathcal{H}_{n}-\mathcal{H}_{4})+2b\left(\tfrac{1}{4}-\tfrac{1}{n}\right)+F_{4}\;.\]
Multiplying by $\tbinom{n}{4}$ and using $\tbinom{n}{4}\cdot\frac{n-4}{n}=\tbinom{n-1}{4}$
gives a telescoping recurrence for $G_{n}:=\tbinom{n}{4}C_{n}$:\begin{align*}
G_{n} & =G_{n-1}\;+\;6a(\mathcal{H}_{n}-\mathcal{H}_{4})\tbinom{n}{4}+2b\left(\tfrac{1}{4}-\tfrac{1}{n}\right)\tbinom{n}{4}+F_{4}\tbinom{n}{4}\\
 & =\sum_{i=5}^{n}\Bigl[6a(\mathcal{H}_{i}-\mathcal{H}_{4})\tbinom{i}{4}+2b\left(\tfrac{1}{4}-\tfrac{1}{i}\right)\tbinom{i}{4}+F_{4}\tbinom{i}{4}\Bigr]+G_{4}\\
 & =\sum_{i=1}^{n}\Bigl[6a(\mathcal{H}_{i}-\mathcal{H}_{4})\tbinom{i}{4}+2b\left(\tfrac{1}{4}-\tfrac{1}{i}\right)\tbinom{i}{4}+F_{4}\tbinom{i}{4}\Bigr]\underbrace{-F_{4}\tbinom{4}{4}+G_{4}}_{=0}\\
 & =6a\sum_{i=1}^{n}\mathcal{H}_{i}\tbinom{i}{4}+(\tfrac{1}{2}b-6\mathcal{H}_{4}a+F_{4})\sum_{i=1}^{n}\tbinom{i}{4}-2b\sum_{i=1}^{n}\tfrac{1}{i}\tbinom{i}{4}\\
 & =6a\tbinom{n+1}{5}\left(\mathcal{H}_{n+1}-\tfrac{1}{5}\right)+(\tfrac{1}{2}b-6\mathcal{H}_{4}a+F_{4})\tbinom{n+1}{5}-2b\sum_{i=1}^{n}\tfrac{1}{4}\tbinom{i-1}{3}\\
 & =6a\tbinom{n+1}{5}\left(\mathcal{H}_{n+1}-\tfrac{1}{5}\right)+(\tfrac{1}{2}b-6\mathcal{H}_{4}a+F_{4})\tbinom{n+1}{5}-\tfrac{1}{2}b\tbinom{n}{4}\;.\end{align*}
Finally, we arrive at an explicit formula for $C_{n}$ valid for $n\ge4$:
\begin{align*}
C_{n} & =G_{n}\bigm/\!\tbinom{n}{4}=\tfrac{6}{5}a\cdot(n+1)\left(\mathcal{H}_{n+1}-\tfrac{1}{5}\right)+(\tfrac{1}{10}b-\tfrac{6}{5}\mathcal{H}_{4}a+\tfrac{1}{5}F_{4})\cdot(n+1)-\tfrac{1}{2}b\;.\end{align*}
Using $F_{4}=5a+b+\frac{1}{2}d$, this simplifies to the claimed closed
form\[
C_{n}=\tfrac{6}{5}a\cdot(n+1)\left(\mathcal{H}_{n+1}-\tfrac{1}{5}\right)+\bigl(-\tfrac{3}{2}a+\tfrac{3}{10}b+\tfrac{1}{10}d\,\bigr)\cdot(n+1)-\tfrac{1}{2}b\;.\]
\pagebreak{}

\section{Explanation for the Curious $\frac{n-q}{n-2}$ Terms\label{sec:Corner-Case}}

All Quicksort variants studied in this paper perform partitioning
by some variant of Hoare's “crossing pointers technique”. This technique
gives rise to two different cases for “crossing”: As the pointers
are moved alternatingly towards each other, one of them will reach
the crossing point \emph{first} -- waiting for the other to arrive.

The asymmetric nature of Algorithm~\ref{alg:Dual-Pivot-Quicksort-Yaroslavskiy}
leads to small differences in the number of swaps and comparisons
in these two cases: If the left pointer $k$ moves last, we always
leave the outer loop of Algorithm~\ref{alg:Dual-Pivot-Quicksort-Yaroslavskiy}
with $k=g+1$ since the loop continues as long as $k\le g$ and $k$
increases by one in each iteration. If $g$ moves last, we decrement
$g$ \strong{and} increment $k$, so we can end up with $k=g+2$.
Consequently, operations that are executed for every value of $k$
experience one additional occurrence.

To precisely analyze the impact of this behavior, the following equivalence
is useful.
\begin{lemma}
\label{lem:Case2}Let $A[1],\dots,A[n]$ contain a random permutation
of $\{1,\dots,n\}$. Then, Algorithm~\ref{alg:Dual-Pivot-Quicksort-Yaroslavskiy}
leaves the outer loop with $k=g+2$ (at line~\ref{lin:yaroslavskiy-end-while})
iff initially $A[q]>q$ holds, where $q=\max\{A[1],A[n]\}$ is the
large pivot.
\end{lemma}
For conciseness, we will abbreviate “Algorithm~\ref{alg:Dual-Pivot-Quicksort-Yaroslavskiy}
leaves the loop with $k=g+i$\,” as “Case $i$\,” for $i=1,2$.
\begin{proof}
Assume Case 2 occurs, i.\,e.~the loop is left with a difference
of 2 between $k$ and $g$. This difference can only show up when
both $k$ is inremented \strong{and} $g$ is decremented. Hence,
in the last iteration we must have entered the else-if-branch in line~\ref{lin:yaroslavskiy-comp-2}
and accordingly $A[k]>q$ must have held there.

Recall that in the end, $q$ is moved to position $g$, so when the
loop is left, at line \ref{lin:yaroslavskiy-end-while} we have $g=q-1$.
By assumption, we are in Case~2, so $k=g+2=q+1$ here. As $k$ has
been increased once since the last test in line~\ref{lin:yaroslavskiy-comp-2},
we know that $A[q]>q$, as claimed.

Assume conversely that $A[q]>q$. As $g$ stops at $q-1$ and is always
decremented in line~\ref{lin:yaroslavskiy-g--}, we have $g=q$ for
the last execution of line~\ref{lin:yaroslavskiy-swap-2}. By assumption
$A[g]=A[q]>q$, so the loop in line~\ref{lin:yaroslavskiy-comp-3}
must have been left because of a violation of condition {}``$k<g$''.
This implies $k\ge g=q$ in line~\ref{lin:yaroslavskiy-swap-2}.
With the following decrement of $g$ and increment of $k$, we leave
the loop with $k\ge g+2$, so we are in Case~2.\qed

Lemma~\ref{lem:Case2} immediately implies that Case~2 occurs with
probability $\frac{n-q}{n-2}$, given pivots $p$ and $q$: For $q<n$,
there are $n-2$ elements that can possibly take position $A[q]$
and $n-q$ of them are $>q$. For $q=n$, we never have $A[q]>q$
and $\frac{n-q}{n-2}=0$.
\end{proof}

\subsection{Additional Contributions to Comparisons}

In Algorithm~\ref{alg:Dual-Pivot-Quicksort-Yaroslavskiy}, the comparison
in line~\ref{lin:yaroslavskiy-comp-1} is executed once for every
value of $k$. Hence, we get an additional contribution of one for
Case~2. For the conditional expectation $c_{n}^{p,q}$, we get an
additional summand $1\cdot\Pr[\mbox{Case 2}]=\frac{n-q}{n-2}$.

Line~\ref{lin:yaroslavskiy-comp-2} is reached for every value of
$k$ with $A[k]\ge p$. By Lemma~\ref{lem:Case2}, Case~2 is equivalent
to $A[q]>q>p$, hence the comparison in line~\ref{lin:yaroslavskiy-comp-2}
is executed exactly once more for $k=q$. This is another contribution
of $\frac{n-q}{n-2}$ to $c_{n}^{p,q}$.

Finally, line~\ref{lin:yaroslavskiy-comp-4} is executed for all
values of $g$ with $A[g]\le q$ \emph{plus} one additional time
in Case~2: As argued in the proof of Lemma~\ref{lem:Case2}, in
Case~2, we always quit the last execution of the loop in line~\ref{lin:yaroslavskiy-comp-3}
because of condition {}``$k<g$'', as the other condition is guaranteed
to hold. Consequently, we get an execution of line~\ref{lin:yaroslavskiy-comp-4}
for $g=q$ even though $A[g]>q$. This comparison is not accounted
for by the terms $\condexpnumberat sG+\condexpnumberat mG$ discussed
in the main text. Hence, it entails an additional contribution of
$\frac{n-q}{n-2}$ for $c_{n}^{p,q}$

The expected number of executions of line~\ref{lin:yaroslavskiy-comp-3},
$|\positionsets G|$, is not affected by Case~2, so no additional
term, here. Summing up, we have $3\cdot\frac{n-q}{n-2}$ additional
comparisons that have not been taken into account by the discussion
in the main text.

\subsection{Additional Contributions to Swaps}

Line~\ref{lin:yaroslavskiy-swap-1} is executed for values of $k$
with $A[k]<p$. In Case~2, $k$ attains one more value, namely $k=q$.
Nevertheless, for this new value of $k$, we do not reach line~\ref{lin:yaroslavskiy-swap-1},
as Lemma~\ref{lem:Case2} tells us that $A[q]>q>p$.

The swap in line~\ref{lin:yaroslavskiy-swap-2} is always followed
by line~\ref{lin:yaroslavskiy-comp-4}, so these lines are visited
equally often. As shown above, line~\ref{lin:yaroslavskiy-comp-4}
causes an additional contribution of $\frac{n-q}{n-2}$.

Finally, the expected number of executions of line~\ref{lin:yaroslavskiy-swap-3},
$\condexpnumberat sG$, is not affected by Case~2, as $\positionsets G$
is the same in Case~1 and~2.

In summary, we find an additional contribution of $\frac{n-q}{n-2}$
to $s_{n}^{p,q}$.
\end{document}